\documentclass{article} 
\pdfoutput=1

\usepackage[authordate,backend=biber]{biblatex-chicago} 
\usepackage{authblk}

\usepackage[english]{babel}
\usepackage[autostyle, english = british]{csquotes}
\usepackage[CJKbookmarks, pdftex, bookmarksnumbered, bookmarksopen, colorlinks, citecolor=blue, linkcolor=blue]{hyperref}
\usepackage{amsmath,amssymb}
\usepackage{graphicx}
\usepackage{enumitem}
\setlist[enumerate]{topsep=0pt,parsep=-1mm,leftmargin=5mm}
\def\be{\begin{equation}}
\def\ee{\end{equation}}

\begin{document}

\title{On the Time Orientation of Probabilistic Theories}

\date{\today}

\author[1]{Andrea {Di Biagio}}
\author[2]{Carlo Rovelli}
\affil[1]{\small Institute for Quantum Optics and Quantum Information (IQOQI) Vienna, Austrian Academy of Sciences, Boltzmanngasse 3, A-1090 Vienna, Austria}
\affil[1]{\small Basic Research Community for Physics e.V., Mariannenstraße 89, Leipzig, Germany}
\affil[2]{\small Aix-Marseille University, Universit\'e de Toulon, CPT-CNRS, F-13288 Marseille, France.}
\affil[2]{\small Department of Philosophy and the Rotman Institute of Philosophy, 1151 Richmond St.~N London  N6A5B7, Canada}
\affil[2]{\small Perimeter Institute, 31 Caroline Street N, Waterloo ON, N2L2Y5, Canada} 
\affil[2]{\small Santa Fe Institute, 1399 Hyde Park Road Santa Fe, New Mexico 87501, USA}

\maketitle
        
\begin{abstract} 
\noindent An influential theorem by Satosi Watanabe convinced many that there can be no physical probabilistic theory with both non-trivial forward and backward transition probabilities. We show that this conclusion does not follow from the theorem. We point out the flaw in the argument, and we showcase examples of theories with well-defined backward and forward transition probabilities.
\end{abstract}

\section{Introduction}

\noindent We often resist discoveries that question the intuitions of everyday experience.  Perhaps no case of such an irrational reaction is as strong as the resistance against the discovery that the fundamental laws of nature do not distinguish past from future, and imply that the vivid orientation of time in our experience is contingent.  

Wrong arguments are commonly invoked to support this resistance.   One of these involves interpreting a theorem by Satosi Watanabe~\parencite[Theorem 3.1]{WS} as a proof that there can be no physical theory specifying both (non-trivial) forward and backward transition probabilities.  Since quantum theory is probabilistic, this interpretation implies that quantum physics is necessarily time oriented. Here we show that this interpretation of the theorem is not correct.

Let $i=1,...,n$ label a set of mutually exclusive events that can happen at a time $t$ and $j=1,...,n$ label a set of mutually exclusive events that can happen at a time $t'$ later than $t$.  Let $p(i,j)$ the probability for both to happen, under suitably specified conditions.   We have immediately the marginal probabilities $p(i)=\sum_j p(i,j)$ for $i$ to happen (at $t$), and  $p(j)=\sum_i p(i,j)$ for $j$ to happen (at $t'$), and the conditional probabilities $p(j|i)=p(i,j)/p(i)$ for $j$ to happen if $i$ happens, and $p(i|j)=p(i,j)/p(j)$ for $i$ to happen if $j$ happens. From the definition, these quantities satisfy the relation
\begin{equation}
p(i|j)\,p(j)=p(j|i)\,p(i).
\label{condition}
\end{equation}

Here is Watanabe's theorem: if $p(i|j)$ and $p(j|i)$ are given, and compatible with this equality, they determine the probabilities $p(i)$ and $p(j)$ \textit{uniquely}. That is: the two conditional probabilities, together, determine the marginals. The proof is simple. It suffices to note that
\begin{equation}\label{proof}
\sum_j\frac{p(j|i)}{p(i|j)}=\sum_j\frac{p(j)}{p(i)}=\frac1{p(i)}\,.
\end{equation}
An interpretation of this result that is widespread in the literature  \parencite{St,healey1981statistical,callender2000time}, and was already suggested in  \parencite{WS}, is that  theories with non-trivial probabilistic state-transition laws are inherently time-asymmetric: a valid physical theory can only specify transition probabilities in one direction of time. The argument is as follows. Suppose a theory specified nontrivial transition probabilities in both directions of time. Then the theory would also specify, via \eqref{proof}, the initial probabilities. But since experimenters are free to fix the probabilities for their experiments, this kind of theory is incompatible with observation.

This argument is wrong, because it mixes probabilities defined by the sole dynamics of the theory with probabilities relevant in a specific  experimental setup. The latter are contextual to the experiment: they depend on how the experiment is set up, for instance on particular conditions selected or chosen. The earlier are not. Below, we show in detail in what sense dynamics alone is sufficient to determine these probabilities.

In classical mechanics, dynamics is given by the equations of motion. No fundamental equation of motion we know distinguishes past from future.   Equations of motions have many solutions, the space of which is (the covariant version of) the phase space of the theory.  This space includes all possible individual  processes permitted by the dynamics.  The actual process, or the statistical mixture of processes, relevant in a given laboratory experiment, or in an observation, is not determined by the theory; it is a \textit{contingent} fact.  It is different in different instances where the theory applies: in different experiments, in different possible worlds satisfying the same dynamics.  The solution that describes a given phenomenon, namely a point in phase space, can be specified in terms of initial values of variables, or final ones, or  values at some intermediate time, or a mixture of these.  Classical mechanics knowns no direction of time. 

It is common to emphasize \emph{initial} values for a contingent reason: we live in a macroscopic world characterized by the second law of thermodynamics and because of this {\em contingent\ } fact, which is not part of the dynamical theory, the present holds macroscopic traces of the past  \parencite{rovelli2022memory}  and macroscopic agents can compute and influence the macroscopic future \parencite{Rovelli2020a,rovelli2022how}.   This naturally leads to thinking in terms of the past `determining' the future, and not the other way around. This logic is appropriate and efficacious macroscopically, but it is not inherent in the mathematics of classical mechanics, whose equations of motion define motions that are not oriented in time. Adding this contingent feature of our world to the mechanical theory, whose mathematics knows nothing about the distinction between past and future, is a conceptual mistake that creates confusion when exploring relativistic and general-covariant physics, where the structure of temporality becomes richer. 

Things are more subtle in a probabilistic theory, and this is what we are concerned with here.   We show how, by keeping the distinction between the theory itself and its contingent applications clear, \emph{it is possible} to have a time reversal invariant probabilistic theory, contrary to claims supported by a wrong interpretation of Watanabe's theorem.  

To be clear, we are not arguing here that time-oriented formulations of statistical or quantum mechanics are \textit{wrong}. Both theories are commonly presented  as time oriented, and there is nothing wrong in this, as formalisms do not need to keep symmetries manifest \parencite{dibiagio2021arrow}. What we are arguing is that time-reversal symmetry, whether manifest or not, underpins classical and quantum physics, and is only broken by the contingent gradient of entropy of our world.  That is, it is wrong to assert that a time-reversal invariant formulation of a probabilistic theories is impossible.   In fact, we show below how statistical mechanics and quantum physics can be naturally formulated in a time-reversal invariant manner. 

\section{Time reversal invariant probabilistic theory}

Before addressing statistical and quantum physics, let's sketch a general form of a time-reversal invariant probabilistic theory.  

A probabilistic theory can be given by assuming that certain processes are more probable and others are less probable.  Let's consider the ensemble of all processes permitted by the theory, and say that the theory assigns a probability to these. In particular, say the theory assignes a probability $p_{t_i,t_j}(i,j)$ to processes where $i$ happens at time $t_i$ and $j$ happens at time $t_j$.
We assume the translation invariance in time, namely, that this probability depends only on the difference $t=t_j-t_i$ and write $p_t(i,j)\equiv p_{t_i,t_j}(i,j)$ if $t={t_j-t_i}$.

We say that this dynamics is invariant under time reversal if
\be
p_{-t}(i,j)=p_t(i,j),
\label{t}
\ee
or, equivalently, 
\be
p_{t}(i,j)=p_{t}(j,i),
\ee
since $p_t(i,j)=p_{-t}(j,i)$ follows from the definitions.  
We are interested in the possibility and coherence of such time reversal-invariant probabilistic theories. For these, \eqref{t} shows that the joint probability depends on the absolute value of time, not its sign, and because of time translation invariance the marginal distribution 
\be
p(i)=\sum_j p_t(i,j)=\sum_j p_t(j,i)
\ee
does not depend on time at all.  Examples below will clarify the physical interpretation of this marginal probability.  The theory then gives the conditional probabilities
\be
p_{t}(i|j)=\frac{p_{t}(j,i)}{p(j)}, \ \ \ \ p_{t}(j|i)=\frac{p_{t}(j,i)}{p(i)}. 
\ee
These are in general different and both depend on $t$ via its absolute value only.

Let's discuss a specific experiment $E$, for instance, in a laboratory at some time.  We  `prepare' an experience or measure some conditional probability in specific realizations.  Because of the time orientation of the macrophysics, we often exploit the agency that this permits and choose or select the {\em initial} value $i$ of an experiment.  If we are confident that there is no other bias, we can immediately use the theory to predict that the probability of $j$ is $p_t(j|i)$. 

More generally, we may be able to set the initial probability of different configurations. Let us call this $p_E(i)$.  We need the subscript $E$ to emphasize the distinction between probabilities {\em in a single experiment} and probabilities $p(i)$ defined purely by the dynamics.  What we do in the lab is to prepare a certain number of runs of the experiment, say $m$ runs, in such a way that a frequency $p_E(i)$, set freely by us, is realized. Letting nature follow its probabilistic ways, we find that the probability distribution of the $j$'s at the later time is going to be given by $p^\mathrm{outcome}_E(j)=\sum_i p_t(j|i)p_E(i)$.   This follows from the above if the cases in the laboratory correctly reflect the distribution of the overall set of data, namely if they are unbiased, except for the arbitrarily chosen selected probability distribution $p_E(i)$.  

Notice that it is then not true that ${p_E(i)=\sum_jp_{t}(i|j)p^\mathrm{outcome}_E(j)}$. There is no reason for this to be true, because the set $p^\mathrm{outcome}_E(j)$ is not an unbiased sample of $p(j)$; it is biased by the choice of $p_E(i)$.

In the special case $p_E(i)=p(i)$, we have immediately $p^\mathrm{outcome}_E(j)=p(j)$, since
\be 
\sum_i p_t(j|i)p(i) =  \sum_i p_t(i,j)=p(j).
\ee
That is, the probability distribution $p(i)$ is stable under time evolution. 

Alternatively, we can make an experiment where we set the {\em final} data. This can be done, for instance, by post-selecting the frequency $p_E(j)$ {\em without any other bias} except for the arbitrarily chosen selected probability distribution $p_E(j)$.  In particular, we should be sure that there is no bias on the initial conditions. In this case, we shall find that $p^\mathrm{outcome}_E(i)=p(i|j)p_E(j)$. 

Both transition probabilities  $p(j|i)$ and $p(i|j)$ are specified by the theory and each correctly governs experimental results.  The one to use depends on the question we ask and the experimental setting we have, which is what can be called `contextuality'.  The context is given by the arbitrarily chosen probabilities $p_E(i)$ or $p_E(j)$.  
Changing the context changes the transition probabilites from the time-symmetric ones.\footnote{In our experience, setting \emph{final} data is more cumbersome and unnatural than setting \emph{initial} data.  The reason are the time orientation of agency and causation. These are themselves rooted in the contingent thermodynamic entropy gradient \parencite{Rovelli2020a,rovelli2022how}. Because of this orientation we read the effect of what we do as affecting the future. This implies that we can  act on initial conditions, without directly affecting final ones, but not viceversa, because to affect the final conditions our agency necessarily disturbs the process itself.}   The transition probabilities $p(j|i)$ describe the phenomena when $p_E(i)$ are chosen freely, while the transition probabilities $p(i|j)$ describe the phenomena when $p_E(j)$ are chosen freely.  There is no contradiction in the fact that they are both specified by the theory and yet we can set initial (or final) contextual probabilities freely.  

There is also a case where they both apply correctly: this is when we make sure there is no bias coming from the past (or future) conditions. In this case it is correct that $p(j|i)$ and $p(i|j)$ together determine all probabilities. We give examples of this in the next two sections.

This is a good time to comment between the distinction between the probabilities and transition probabilities given by a theory, and the frequencies observed in the world, within or without experimental settings. If we look around and record the frequency of the occurrence of various phenomena and these do not agree with the frequencies in the theory, can we conclude that the theory is wrong? This is not the case. Consider, for the moment, the case of classical theory. The theory comes equipped with a phase space which represents all the possible histories of a system. Should we expect all these histories to be realised in nature? No. Some histories may never occur. The role of the theory, in other words, is not to tell us what happens in the world, but to allow us to say what happens in the world \textit{given} what other things happen. The probabilistic theory, then, tells us how to compute the probability of various events given information about other---past \textit{and/or} future---events, and their context.

In this section, we have shown how a theory can specify nontrivial backwards and forwards probability distributions. In the following, we show how statistical and quantum mechanics can be set in this form and we discuss the physical interpretation of the probability distribution $p(i)$ in these cases.

\section{Quantum theory}

Consider a quantum system with Hilbert space $\cal H$, time independent Hamiltonian $H$ and time reversal operator $T:{\cal H}\to{\cal H}$. The evolution operator $U(t)=e^{-i H t/\hbar}$ satisfies $TU(t)=U(-t)T$.  In the Schr\"odinger basis, ${\cal H}$ is formed by complex wave functions $\psi$ on configuration space, and $(T\psi)(x)=\overline{\psi(x)}$.  In general, the operator $T$ is antinunitary and satistfies $T^2=1\!\!1$, because reversing time twice has no effect. The position operator is even, namely $xT=Tx$. 
The dynamics of the theory can be given by the probabilities 
\be
p_t(\psi|\phi)=| \langle \psi |e^{-\frac{i}{\hbar}Ht}| \phi \rangle |^2. 
\ee
Let $|i\rangle$ be a basis.  Let us first consider the case of finite dimensional Hilbert spaces with dimension $d$, for simplicity.   
The transition probabilities are satisfy 
\begin{equation}
p_t(i|j)
=| \langle i |U(t)| j \rangle |^2 
= | \langle j |U(-t)| i \rangle |^2  = p_{-t}(j|i). 
\end{equation}
Using time reversal invariance, namely ${U(-t)=TU(t)T}$, we also have 
\begin{equation}
p_{-t}(i|j)
=| \langle i |U(-t)| j \rangle |^2 
= | \langle i |TU(t)T| j \rangle |^2  = p_{t}(Ti|Tj).
\end{equation}
If the basis $|i\rangle$ diagonalizes an even operator, 
\begin{eqnarray}
p_{-t}(i|j) = p_{t}(i|j) = p_{t}(j|i)  . 
\end{eqnarray}
It follows that 
\be
\frac{1}{p(i)}=\sum_j\frac{p_t(i|j)}{p_{t}(j|i)}=d,
\ee
and thus
\be
p(i)=p(j)=\frac1d.
\label{pd}
\ee
That is, there is a natural analog to the micro-canonical equilibrium distribution considered in the previous section.  In the previous case we first defined the time symmetric theory via joint probabilities. Here, rather, we have observed that the forwards and backwards transition probabilities do so for quantum theory. 

Let us give a physical interpretation to the probability distribution \eqref{pd}.  Suppose we prepare a generic density state $\rho_o$, this undergoes a sequence of measurements of non commuting observables (whose results we do not know) and we are interested in the probabilistic state after these measurements.   For a sufficiently long sequence, $\rho_o$ will evolve to the completely spread density ${\rho=\sum_i|i\rangle\!\langle i| /d=1\!\!1/d}$, because at every measurement some information about the initial probability distribution is lost.  For instance, a qubit in any probabilistic mixture of eigenstates of the $z$ basis has probability 1/2 of being in any element of the $y$ basis after a single $y$ measurement.

It is immediate to see that these quantities satisfy \eqref{condition}.  This definition of probabilities treats any $i$ and any $j$ both as independently a priori equiprobable, but correlated. Any specific selection of $p_E(i)$ determines some $p_E^\mathrm{outcome}(j)$ and viceversa, as for the formulas above. Transition probabilities are well defined and each is physically meaningful in its appropriate context. 

The above argument holds for also for systems with infinite-dimensional Hilbert spaces, whenever one considers finite-dimensional energy eigenspaces. We also note that one can extend this beyond pure preparations and measurements on orthonormal bases, and indeed to post-quantum operational theories \parencite{dibiagio2021arrow,hardy2021time,selby2022time}.

As a concrete example, consider the simple case where $i=\pm$ indicates that a qubit has been measured to have spin $\pm$ in the $z$ direction at $t$ and $j=\pm$ indicates that it has been measured to have spin $\pm$ in a direction $r$ at angle $\phi$ from the $z$ direction at $t'$, and assume a vanishing hamiltonian. Quantum theory predicts
\be
p(j|i)=p(i|j)=\begin{pmatrix}
\cos^2(\phi/2)&\sin^2(\phi/2) \\
\sin^2(\phi/2)&\cos^2(\phi/2)
\end{pmatrix}.
\label{qm}
\ee
To measure $p(j|i)$ we can select the particles with spin $i$ in the $z$ direction at time $t$ out of an unbiased set, and then measure the spin in the $r$ direction at $t'$. The fraction of outcomes is $p(j|i)$.  To measure $p(i|j)$ we can select the particles measured as having with spin $j$ in the $r$ direction at time $t'$ {\em out of an unbiased set}, and count the fraction of these whose spin was measured at $t$ to be $i$ in the $z$ direction. The fraction of outcomes is $p(i|j)$.   

To concretely realize unbiased sets we can for instance proceeded as follows.  We measure the spin of the particle four times, in sequence, as follows. (a): in a direction orthogonal to the $z$ axis, (b): in the direction of the $z$ axis, (c): in the direction of the $r$ axis, (d): in a direction orthogonal to the $r$ axis.   Alternatively, we can reverse the time order of the sequence, without any effect on what follows.  The (a) and (d) measurements have the effect of fully un-biasing the two outcomes of the (b) and (c) measurements, because the probability distribution of (b) outcomes given any (a) outcome, and  the probability distribution of (c) outcomes given any (d) outcome, are flat.  Then we can take the frequency of $(i,j)$ outcomes of the two measurements (b) and (c) as a measure of $p(i,j)$, and verify that \eqref{qm} gives the correct prediction. In the last section, we will comment further on the fact that, in practice, there is no need to perform the unbiasing measures (a) and (d) if we want to measure the transition probabilities from past to future.

\section{Classical Ergodic Dynamics}

Consider a classical Hamiltonian system with time independent Hamiltonian. The equations of motion define a flow both in configuration space $\cal C$ (the space of generalised coordinates) and phase space $\Gamma$ (the space of generalised coordinates and momenta).

We say that the system is \textit{ergodic} if there is a probability distribution $p$ on a subspace\footnote{To be clear, $\gamma$ will in general be a proper subspace of $\Gamma,$ lying at the intersection of level surfaces of constants of the motion such as energy, angular momentum, etc.} $\gamma\subset\Gamma$ such that $p$ is the limit of any probability distribution $p_o$ on $\Gamma$, that is if $\lim_{t\to-\infty}p_o(s_{t})=p(s)$ for all $s$ in $\gamma$, where $s_t$ is the point that $s$ is mapped to by the dynamics in time $t$. The \textit{equilibrium distribution} $p$ is entirely determined by the dynamics \parencite{Gibbs,Wayne}. By time-reversal invariance, it is also true that $p(s)=\lim_{t\rightarrow\infty}p_o(s_{t})$, and by time-translation invariance we also have $p(s)=p(s_t)$.

We can also see the equilibrium distribution $p$ as a distribution over \textit{solutions} $h:\mathbb{R}\to\Gamma$ to the equations of motion. This is done by recalling that the phase space $\Gamma$ can be thought of as the space of solutions by identifying a solution with its conditions at a fixed reference time $t_0$, that is, by identifying a point $s\in\Gamma$ with the solution $h_s$ such that $h_s(t_0)=s.$

If $h: t\mapsto (x(t),\pi(t))$ is a physical motion in phase space, then its time-reversal is $h^{-1}:t\mapsto(x(-t),-\pi(-t))=Th(-t),$ where we introduced the map $T:\Gamma\to\Gamma$ that inverts momenta.  The map $T$ can be extended to observables $F:\Gamma\to\mathbb R$ by defining a new observable $TF(s)=F(Ts)$. We say that an observable $F$ is even if $TF=F$ or odd if $TF=-F.$

Consider two observables $F,G:\Gamma\rightarrow\mathbb R$, and denote by $\{f,t_\mathrm{i},g,t_\mathrm{f}\}\subset \Gamma$ the subset of solutions $h$ such that $F(h(t_\mathrm{i}))=f$ and $G(h(t_\mathrm{f}))=g.$ That is, all motions such that $F$ has value $f$ at $t_\mathrm{i}$ and $G$ has value $g$ at $t_\mathrm{f}$. Define the joint probability
\begin{equation}
    p_{t_\mathrm{i},t_\mathrm{f}}(f,g)=\int_{t_\mathrm{i},t_\mathrm{f}} p.
\end{equation}
Since $\{f,t_\mathrm{i},g,t_\mathrm{f}\}=\{g,t_\mathrm{f},f,t_\mathrm{i}\}$ by definition, we have
\begin{equation}
    p_{t_\mathrm{i},t_\mathrm{f}}(f,g) = p_{t_\mathrm{f},t_\mathrm{i}}(g,f).
\end{equation}
By time-translation invariance, the probability depends only on the linterval $t=t_\mathrm{f}-t_\mathrm{i},$ so we can write
\begin{equation}
    p_t(f,g)=p_{t_\mathrm{i},t_\mathrm{f}}(f,g),
\end{equation}
and the previous equation reads
\begin{equation}\label{flip}
    p_t(f,g) = p_{-t}(g,f).
\end{equation}

Recalling that $h$ is a solution if and only if its time-reversal is $h^{-1}$ a solution establishes a bijection between $\{f,t_\mathrm{i},g,t_\mathrm{f}\}$ and $\{Tf,-t_\mathrm{i},Tg,-t_\mathrm{f}\}$. Then, since by time-reversal invariance of $p$, 
\begin{equation}
    p(h)=p(h^{-1}),
\end{equation}
therefore we have that
\begin{equation}
    p_t(Tg,Tf)=p_t(f,g).
\end{equation}
Using \eqref{flip}, this reads
\begin{equation}
    p_t(Tg,Tf) = p_{-t}(g,f).
\end{equation}
Focussing on the case where $F$ and $G$ are both even observables we obtain the two equations
\be
p_{-t}(f,g)=p_t(f,g). 
\ee
and
\be
p_t(f,g)=p_t(g,f). 
\ee
This means that the joint probability of having two given values of two even observables does not depend on which comes first.

We can then define the probability  
\be
p(f) = \int_{\{f,t\}} p,
\ee
which is independent from time.  We could now consider the entropy  
\be
S(f)= S_0+k\ln p(f),
\ee
where $S_0$ contains the normalisation of the phase-space volume (and, with input from quantum mechanics, the volume of the smallest available cell) and we have inserted the Boltzmann constant $k$ to give $S$ the conventional dimensions. 

We are now exactly in the setting of the previous section. In this formulation, both forward and backward conditional probabilities are well defined, contrary to claims based on the wrong interpretation of Watanabe's theorem.  The conditional probability 
\be
p_t(f|g)\equiv \frac{p_t(f,g)}{p(g)}
\ee
gives the probabilistic evolution of the system ahead in time, and
\be
p_t(g|f)\equiv \frac{p_t(f,g)}{p(f)}
\ee
back in time.

Note that these two conditional probabilities will in general be different.
This is because the probability for going to higher entropy  is larger than the probability of going to lower entropy.  But both depend on time via its absolute value only. 

At this point it is clear what is the meaning of the distribution $p(f)$ in this context: it is the equilibrium distribution. It is invariant under the evolution defined by the conditional probabilities:
\be
p(f)=\int dg\ p_t(f|g)\; p(g).
\ee 
and is the limit probability in the sense that 
\be
p(f)=\lim_{t\to\infty}\int dg\ p_t(f|g)\; p_o(g).
\ee 
for any $p_o$.  In fact, this evolution increases the entropy, which is maximized (on $\gamma$) by $p$.

The equilibrium distribution is the usual micro-canonical distribution. Notice that this is not chosen as an independent postulate for statistical mechanics: it is determined by the (ergodic) dynamics.  The time-reversal invariant probability distribution $p_t(f,g)$ describes equilibrium, where there is no preferred time orientation.   Any specific experiment, even away from equilibrium, is accounted for by the theory by simply taking the contingent probabilities $p_E(f)$ into account.    What breaks time reversal invariance in our world is the environmental entropy gradient, which makes it natural for us to think in terms of initial conditions determining the future.

\section{The time orientation of our world.}

We have shown that Watanabe's theorem does not imply that theories with non-trivial probabilistic state-transition laws are necessarily time-asymmetric.  The reason we judge it is important to point this out is that this wrong interpretation of the theorem mistakenly precludes a possible coherent interpretation of the ubiquitous time orientation of our world as due to contingency rather than dynamics. That is, as due to the particular solution of the dynamics in which we live, rather than the dynamical laws themselves.

The concrete procedures to set unbiased final or initial conditions are not symmetric in a realistic laboratory. In a lab, we do not need to do anything special to be sure that that the \emph{later} probabilities are unbiased, while we do have to do something special to be sure that that the {\em earlier} probabilities are unbiased. In the example above, we actually need only \textit{one} of (a) and (d) if we want to observe the \textit{backward} transition probabilities and \textit{neither} if we want to observe \textit{forward} transition probabilities.  But this fact does not necessarily imply that  the dynamical laws of nature are time oriented. Rather, it may simply be due to the fact that we live in a (contingent) entropy gradient that has the effect that we have traces of past and not analogous traces of the future \parencite{rovelli2022memory} and we have time-oriented agency that affects the future and not the past  \parencite{Rovelli2020a,rovelli2022how}. The effect of this entropy gradient is that if $i$ comes earlier than $j$, it is automatic to screen $j$ from any other bias besides the one determined by $i$: it suffices to act earlier than $i$ to set $i$ appropriately. Then the probability distribution of $j$ is biased only by $i$ in the way determined by the dynamics.  But it is hard to set $j$ instead, without disturbing $i$, because to do so without interfering with the process itself, we can only act earlier than $i$, thus affecting $i$, which is in between. The asymmetry can therefore simply be the effect of the contingent entropy gradient, a feature of a specific system in the theory---our world. It may be not a fundamental asymmetry in the laws of the theory, which in itself specifies transition probabilities in both directions of time.

We have seen that a theory with nontrivial backwards and forwards transition probabilities can exist, and this is compatible with our ability to freely set the probabilities of specific events.  The key insight is that, in intervening to change the probabilities in a specific experiment, we are modifying the context of the phenomenon in such a way that the original probabilities of the theory, those of the undisturbed system, are modified. This is not a radical thing: in any experiment, one must make sure that the setup is such that it minimises extraneous influences, so that the theoretical model one uses reflects the experiment. The time-asymmetric effect of our interventions on a system is a contingent fact of our world, not built in the dynamical laws.

\section*{Acknowledgements}
We acknowledge support of the ID~\#~61466 grant from the John Templeton Foundation, as part of the ``Quantum Information Structure of Spacetime (QISS)'' project (\hyperlink{http://www.qiss.fr}{qiss.fr}).


\printbibliography

\end{document}